\documentclass[aps,prl,twocolumn,showpacs]{revtex4}

\begin{document}


\title{Entropy and energy of a class of spacetimes  with  horizon: a general derivation}


\author{T.~Padmanabhan}
\affiliation{IUCAA, 
Post Bag 4, Ganeshkhind, Pune - 411 007\\
email: nabhan@iucaa.ernet.in}


\date{\today}

\begin{abstract}
Euclidean continuation of several Lorentzian  spacetimes with horizons requires treating the
Euclidean time coordinate to be periodic with some period $\beta$. Such spacetimes (Schwarzschild, deSitter,Rindler .....) allow a  temperature $T=\beta^{-1}$ to be associated with the horizon. I construct a canonical
ensemble of a subclass of such spacetimes with a fixed value for $\beta$ and evaluate the partition function $Z(\beta)$. For  spherically symmetric spacetimes with a horizon at $r=a$,  the  partition function has the generic form
$Z\propto \exp[S-\beta E]$, where $S= (1/4) 4\pi a^2$  and $|E|=(a/2)$.  Both $S$ and $E$ are determined entirely by the properties of the metric near the horizon. This analysis reproduces the conventional result for the blackhole spacetimes and provides a simple and consistent interpretation of entropy and energy for deSitter spacetime. For the Rindler spacetime the entropy per unit transverse area turns out to be $(1/4)$ while the energy is zero. The implications are discussed. 
\end{abstract}


\maketitle

\noindent
Among the class of Lorentzian spacetime metrics which allow a positive definite continuation to the Euclidian time coordinate $\tau=it$, there exists a subclass of spacetime metrics which require $\tau$ to be treated a periodic with some period $\beta$. This usually leads to two-point functions of the quantum field theory, defined via Euclidian continuation, to satisfy the KMS condition. It is natural to interpret such a feature as describing
a    finite temperature field theory with temperature $T=\beta^{-1}$. 
(For a review, see e.g., [1].)
A wide class of such spacetimes, analysed in the literature, has the form
    \begin{equation}
     ds^2=f(r)dt^2-f(r)^{-1}dr^2 -dL_\perp^2
     \label{basemetric}
     \end{equation}
  where $f(r)$ vanishes at some surface $r=a$, say, with $f'(a)\equiv B$ remaining finite. When $dL_\perp^2$
  is taken as the metric on 2-sphere and $r$ is interpreted as the radial coordinate $[0\leq r\leq \infty]$, equation (\ref{basemetric}) covers a variety of spherically symmetric spacetimes  (including Schwarzschild, Reissner-Nordstrom, deSitter etc.) with a compact horizon at $r=a$. If $r$ is interpreted as one of the cartesian coordinates $x$ with $(-\infty\leq x\leq \infty)$ and $dL_\perp^2=dy^2+dz^2, f(x)=1+2gx,$ equation (\ref{basemetric}) can describe the Rindler frame in flat spacetime. We shall first concentrate on compact horizons with $r$ interpreted as radial coordinate,
and comment on the Rindler frame at the end.

 Since the metric is static, Euclidean continuation is trivially effected by $t\to
  \tau=it$ and an examination of the conical singularity near $r=a$ [where $f(r) \approx B(r-a)$] shows that $\tau$ should be interpreted as periodic with period $\beta=4\pi/|B|$ corresponding to the temperature $T=|B|/4\pi$. One can prove quite rigorously $^{[1,2]}$ that the spacetime described by (\ref{basemetric}) is endowed with this temperature which
  --- in turn --- depends
  {\it only on the behaviour of the metric near the horizon}. The form of $f(r)$ is arbitrary except for the constraint that $f(r)\approx B (r-a)$ near $r=a$.
  
  The next logical question will be whether one can associate other thermodynamic quantities, especially the entropy, with such spacetimes.$^3$ Given that the temperature can be introduced very naturally, just using the behaviour of metric near the horizon, one would look for a similarly elegant and natural derivation of the entropy.
   Such a derivation should not depend on the introduction of external degrees of freedom (like a scalar field) since we want to associate the entropy with the spacetime and not with an external field. Further,  the thermodynamical description should depend only on the behaviour of the metric near the horizon. I will  show that 
  it  is indeed possible to provide such a description  for spacetimes of the form in (\ref{basemetric}), in spite of the fact that notion of energy is ill defined in a generic spacetime. 
  
 The class of metrics in
  (\ref{basemetric}) with the behaviour $[f(a)=0,f'(a)=B]$ constitute a canonical ensemble at constant temperature  since they all have the same temperature $T=|B|/4\pi$ . The partition function for this ensemble is given  by the path integral sum
   \begin{eqnarray}
    Z(\beta)&=&\sum_{g\epsilon {\cal S}}\exp (-A_E(g))  \\
 &=&\sum_{g\epsilon {\cal S}}\exp \left(-{1\over 16\pi}\int_0^\beta  d\tau \int d^3x \sqrt{g_E}R_E[f(r)]\right)\nonumber
     \label{zdef}
     \end{eqnarray}
  where I have made the Euclidian continuation of the Einstein action and imposed the periodicity in $\tau$
  with period $\beta=4\pi/|B|$.
  The sum is restricted to the set ${\cal S}$ of all metrics of the form in 
  (\ref{basemetric}) with the behaviour $[f(a)=0,f'(a)=B]$ and the Euclidean lagrangian is a functional of $f(r)$.
  No source term or cosmological constant (which cannot be distinguished from certain form of source) is included since the idea is to obtain a result which depends purely on the geometry.
   The spatial integration will be restricted to a region bounded by the 2-spheres $r=a$ and $r=b$, where
  the choice of $b$ is arbitrary except for the requirement that  within the region of integration the Lorentzian
  metric must have the proper signature with $t$ being a time coordinate. The remarkable feature is the form of the
  Euclidean action for this class of spacetimes.  Using the result
  \begin{equation}
  R=\nabla_r^2 f -{2\over r^2}{d\over dr}\left[r(1-f)\right]\label{reqn}
  \end{equation}
  valid for metrics of the form in (\ref{basemetric}), a
   straight forward calculation shows that
   \begin{eqnarray}
  - A_E&=&{\beta\over 4}\int_a^b dr\left[-[r^2f']'+2[r(1-f)]'\right] \nonumber\\
&=&{\beta\over 4}[a^2B -2a]+Q[f(b),f'(b)]
     \label{zres}
     \end{eqnarray}
where $Q$ depends on the behaviour of the metric near $r=b$ and we have used the 
conditions $[f(a)=0,f'(a)=B]$. The sum in (\ref{zdef}) now reduces to summing over the values of $[f(b),f'(b)]$
with a suitable (but unknown) measure. This sum, however, will only lead to a factor which we can ignore in deciding about the dependence of $Z(\beta)$ on the form of the metric near $r=a$. Using $\beta=4\pi/B$
(and taking $B>0$, for the moment)  the final result can be written in a very suggestive form:
 \begin{equation}
  Z(\beta)=Z_0\exp \left[{1\over 4}(4\pi a^2) -\beta({a\over 2})  \right]\propto 
  \exp \left[S(a) -\beta E(a)  \right]
     \label{zresone}
     \end{equation}
with the identifications for the entropy and energy being given by:
\begin{equation}
S={1\over 4} (4\pi a^2) = {1\over 4} A_{\rm horizon}; \quad E = {1\over 2} a = \left( {A_{\rm horizon}\over 16 \pi}\right)^{1/2}
\end{equation}
  In addition to the simplicity, 
     the following features are noteworthy regarding this result:
     
    (i) The result is local in the sense that it depends only on the form of the metric near the 
    horizon.  In particular, the definition of energy does not depend on the asymptotic flatness of the metric.
    
    (ii) The partition function was evaluated with two very natural conditions: $f(a) =0$ making 
    the surface $r=a$ a compact horizon and $f'(a) = $ constant which is the 
    proper characterisation of the canonical ensemble of spacetime metrics. Since temperature
    is well defined for the class of metrics which I have considered, this canonical ensemble is defined without any   ambiguity.  This allows me to  sum over a  class of spherically symmetric spacetimes at one go rather than
    deal with, say, blackhole spacetimes and deSitter spacetime separately. Unlike many of the 
previous approaches,   I do {\it not} evaluate the
path integral in the WKB limit, confining to metrics which are solutions of Einstein's equations. 
 (When the path integral sum is evaluated  using WKB ansatz for vacuum spacetimes like Schwarzschild blackhole 
--- as, e.g., in the  works by Gibbons and Hawking$^4$ --- the scalar curvature
$R$ vanishes and a surface contribution arises from the trace
 of the second fundamental form on the boundary. 
The surface contribution which arises in (\ref{zres}) is different.) Conceptually,
a canonical ensemble for a minisuperspace of metrics  of the form in (\ref{basemetric})  should 
be constructed by keeping the temperature constant {\it without} assuming the metrics
to be the solutions of Einstein's equation; this is what I do and exploit the form of $R$ given by (\ref{reqn}).
Since this action involves second derivatives, it is not only allowed but even required to fix both $f$ and $f'$
at the boundaries.

    (iii) In the case of the Schwarzschild blackhole with $a=2M$, the energy turns out to
    be $E=(a/2) = M$ which is as expected. (More generally,
	$E=(A_{\rm horizon}/16\pi)^{1/2}$ corresponds to the so called `irreducible mass' in 
       BH spacetimes$^5$). Of course, the identifications $S=(4\pi M^2)$,
    $E=M$, $T=(1/8\pi M)$ are consistent with the result $dE = TdS$ in this particular case.
    The result, however, is applicable to a much wider 
     class of spacetimes. Consider, for example,  the class of all spacetimes for which $f(r)=0$ at 
    $r=2M$ but is widely different   from the Schwarzschild metric  (but well behaved) for $r\gg 2M$. Such a 
    class will include spacetimes which are not asymptotically flat so that no natural definition of energy
    exists in the conventional sense. The analysis above suggests that the energy of such spacetimes
    for the purpose of thermodynamics is still given by $E=M$. 
    
    (iv) Most importantly, our analysis provides an interpretation of entropy and energy in the case
    of deSitter universe which is gaining in popularity. In this case, $f(r) = (1-H^2r^2)$, $a=H^{-1}, B=-2H$.
Since the region where $t$ is timelike is ``inside'' the horizon, the integral for $A_E$ in (\ref{zres}) should be taken from some arbitrary value $r=b$ to $r=a$ with $a>b$. So the horizon contributes in the upper limit of the integral
introducing a change of sign in (\ref{zres}). Further, since $B<0$, there is another negative sign in the area term
from $\beta B\propto B/|B|$. Taking all these into account we get, in this case, 
\begin{equation}
  Z(\beta)=Z_0\exp \left[{1\over 4}(4\pi a^2) +\beta({a\over 2})  \right]\propto 
  \exp \left[S(a) -\beta E(a)  \right]
     \label{zrestwo}
     \end{equation}
giving
$S=(1/ 4) (4\pi a^2) = (1/ 4) A_{\rm horizon}$ and  $E=-(1/2)H^{-1}$. 
These definitions do satisfy the relation $TdS -PdV =dE$ when it is noted that the deSitter universe has 
    a non zero pressure $P=-\rho_\Lambda=-E/V$ associated with the cosmological constant. In fact,
if we use the ``reasonable" assumptions $S=(1/4)(4\pi H^{-2}), V=(4\pi/3)H^{-3}$ and $E=-PV$ in the equation
$TdS -PdV =dE$ and treat $E$ as an unknown function of $H$, we get the equation $H^2(dE/dH)=-(3EH+1)$
which integrates to give precisely $E=-(1/2)H^{-1}. $ This energy is also
 numerically same as the
total energy within the Hubble volume of the classical solution, with a cosmological constant:
    \begin{equation}
    E_{\rm Hub}={4\pi \over 3} H^{-3} \rho_\Lambda = {4\pi \over 3} H^{-3} {3H^2\over 8\pi } ={1\over 2}H^{-1}
    \end{equation}
  (The extra negative sign of $E=-E_{\rm Hub}$ is related to a feature noticed in the literature
in a different context; see for example, the discussion following equation (71) in the review [6]. )

 Let us now consider the spacetimes with planar symmetry for which (\ref{basemetric}) is still applicable with $r=x$ being a Cartesian coordinate and $dL_\perp^2=dy^2+dz^2$. In this case $R=f''(x)$ and the action becomes
\begin{eqnarray}
-A_E&=&{1\over 16\pi}\int_0^\beta d\tau\int dy dz \int_a^b dx f''(x)\nonumber\\
&=& {\beta\over 16\pi} A_\perp f'(a)+Q[f'(b)]
\label{rindleraction}
\end{eqnarray}
where we have confined the transverse integrations to a surface of area $A_\perp$. If we now sum over
all the metrics with $f(a)=0,f'(a)=B$ and $f'(b)$ arbitrary, the partition function will become
\begin{equation}
Z(\beta)=Z_0\exp({1\over 4}A_\perp)
\end{equation}
which suggests that planar horizons have an entropy of (1/4) per unit transverse area but zero energy. This includes
Rindler frame as a special case. Note that if we freeze  $f$ to its Rindler form $f=1+2gx$, 
(by demanding the validity of Einstein's equations in the WKB approach, say)
then $R=f''=0$ as it should.
In the action in (\ref{rindleraction}), $f'(a)-f'(b)$ will give zero. It is only because I am
 {\it not} doing a WKB analysis --- but 
 varying $f'(b)$ with fixed $f'(a)$ --- that I obtain an entropy for these spacetimes.  
    
    I shall now indicate how this analysis can possibly be generalized to handle a wider class of 
    spacetimes with compact horizons.  Consider any static metric which requires periodicity
    in $\tau$ under Euclidean continuation. In evaluating the Euclidean action, the integral over
    $\tau$ just leads to a factor $\beta$. The spatial integration is over a region bounded by
    the horizon ${\cal H}$  and another arbitrary, convenient compact 2-surface which does not
    intersect the horizon. Since the Lagrangian density $\sqrt{-g} R$ contains a total
    derivative term
    \begin{equation}
    \sqrt{-g} R = -\partial_l \left[ \partial_a \left( \sqrt{-g}\, g^{al}\right) +  g^{il}  
    \partial_i (\sqrt{-g})\right] + (\Gamma\,  \Gamma \ {\rm terms})
    \end{equation}
    the Euclidean action will pick up two surface  terms on the horizon: 
    \begin{eqnarray}
    A_E &=&  - {\beta\over 16\pi} \int_{{\cal H}} d^2 \sigma \, n_b \partial_a  \left( \sqrt{-g}\, g^{ab}\right) \nonumber\\
 &-&      {\beta\over 16\pi}  \int_{{\cal H}} d^2 \sigma \, n_l \, g^{il}   \partial_i (\sqrt{-g})
    + (\Gamma  \Gamma \ {\rm terms})
    \end{eqnarray} 
    where $n_a$ is the normal to the surface.
    Our analysis suggests that the first term corresponds to the entropy $S$ 
     when ${\cal H}$ is a compact horizon.
     The contribution to $\beta E$ arises from the remaining terms. (In (\ref{reqn}), for example, the first term,
      $\nabla_r^2 f$, arises from the embedding $t=$ constant space in 4-dimensions while the second
term $2r^{-2}[r(1-f)]'$ is the scalar curvature, ${}^3R$, of the $t=$ constant surface).

Another issue which is not adequetely addressed in the literature is the definition of entropy and temperature in spacetimes with more than one horizon --- like the Schwarzchild-deSitter solution. Since the surface gravities at the two horizons are different, one would expect any local definition to lead to two different temperatures and peridicity arguments become ambiguous. Our analysis uses a canonical ensemble which, in turn, assumes that there is {\it single} temperature associated with the spacetime. But if we ignore this feature and formally extend the result in  (\ref{zres}) to
a region between two horizons, we will get the sum $\sum_i(S_i-\beta_i E_i)$ with $i=1,2$.  While summing over the
entropy and energy seems reasonable --- they being extensive --- it is difficult to interpret the occurrence of two different $\beta$'s in a canonical ensemble. These difficulties, of course, have nothing to do the approach described here and exist in all other approaches as well. 

Finally, it will be straight forward to use this approach in $D$ dimensions with the hope that insights gained in $D\neq 4$ may be of some help. In $D=(1+2)$, for example, metrics of the type in (\ref{basemetric}) with $dL^2_\perp=r^2 d\theta^2$ will give $S=(1/4)(2\pi a)=(1/4)A_{\rm horizon}$ with $E=0$. The vanishing of energy probably signifies the fact that at the level of the metric, Einstein's equations are vacuous in (1+2) and we have not incorporated any topological
effects [like deficit angles corresponding to point masses in (1+2) dimensions] in our approach.
 These issues are 
under study. 

I thank S.Shankaranarayanan, K.Subramanian and Suneeta Varadarajan for useful discussions.


\end{document}